# *Polytopes and Nuclear Structure*

*by*

*Roger Ellman*

Abstract

While the parameters $Z$ and $A$ indicate a general structural pattern for the atomic nuclei, the exact nuclear masses in their fine differences appear to vary somewhat randomly, seem not to exhibit the orderly kind of logical system that nature must exhibit.

It is shown that separation energy [the mass of the nucleus before decay less the mass of a decay's products], not mass defect [the sum of the nuclear protons and neutrons masses less the nuclear mass], is the "touchstone" of nuclear stability.

When not part of an atomic nucleus the neutron decays into a proton and an electron indicating that it can be deemed a combination of those two. Considering that, the nucleus can be analyzed as an assembly of $A$ protons and $N = A - Z$ electrons, where $N$ of the protons form neutrons with the $N$ electrons.

Resulting analysis discloses a comprehensive orderly structure among the actual nuclear masses of all the nuclear types and isotopes. The analysis examines in detail the conditions for nuclear stability / instability. An interesting secondary component of that analysis and the resulting logical order is the family of geometric forms called polytopes, in particular the regular polyhedrons.

Roger Ellman, The-Origin Foundation, Inc.
http://www.The-Origin.org
320 Gemma Circle, Santa Rosa, CA 95404, USA
RogerEllman@The-Origin.org

# *Polytopes and Nuclear Structure*

*Roger Ellman*

While the parameters, $Z$ and $A$, of atomic nuclei indicate a general structural pattern for the nuclei, the exact nuclear masses in their various differences seem not to exhibit the orderly kind of logical system that nature must exhibit.

At first consideration atomic nuclei are considered as an assembly of $Z$ protons and $N = A - Z$ neutrons. That description is not to say that such an assembling action actually occurs as such. Rather, the assembly point of view is a procedure for determining what the characteristics of the resulting nucleus must be: it must have a mass defect relative to the sum of the masses of those components in an amount equal to that which is required by the theoretical scenario of so assembling the nucleus.

That is, to assemble the particles as in a nucleus and make them stay so assembled requires removing from them the potential energy that they would have when assembled were it not somehow removed.

Unlike the case of the neutron as a combination of a single proton and electron, the components of an atomic nucleus cannot come together to form the nucleus naturally and unaided because of the mutual electrostatic repulsion of the protons and the electric neutrality of the neutrons.

There are only two ways in practical reality that such a nucleus can come into existence. One is through the radioactive decay of a more complex nucleus. The other is for some two less complex nuclei to be accelerated toward each other with so much energy that they merge in spite of their mutual repulsion. At the moment of merger not only would the new nucleus be formed, but in addition the excess mass would be given off in some combination of small particles and photons.

However, analysis of the nuclear structure as if an assembly of proton and neutron components could be made to occur produces significant further understanding of the nuclear structure. Furthermore, considering that the neutron can be a combination of a proton and an electron, the nucleus can be analyzed as an assembly of $A$ protons and $N = A - Z$ electrons, where $N$ of the protons form neutrons with the $N$ electrons.

*Separation energy (SE)* is the measure of nuclear stability because it deals with the possibility of decay of a nuclear type. It is the mass of the nucleus before decay less the mass of the decay products as in equation *(1)*.

```
(1)  Separation  =  Mass of nucleus
     Energy         before decay

                  + One electron mass if the decay is
                    by the nucleus capturing an electron

                  - Mass of resulting nucleus

                  - Mass of particle(s) emitted
```

If the separation energy is positive then the initial component(s) have enough mass to make up the final components plus some extra mass to appear as energy of motion of the final components or as E-M radiation. If the *SE* is negative then the decay cannot take place because there is not enough mass to make up the final components and conservation would be violated. Therefore, positive *SE* means instability and negative *SE* means stability.

*The 1983 Atomic Mass Evaluation* by The National Institute of Nuclear Physics and High-Energy Physics, Amsterdam; University of Technology, Delft, The Netherlands; and Laboratoire Rene Bernas du CSNSM, Orsay, France, provides the measured data on the nuclear types. Table 1, below, presents a few examples of that data.

*Table 1*
*Some Natural Atomic Types and Masses*

| | Z | A | Measured Atomic Mass amu | Emission if any | Mass Defic'y µ-amu | Separ'n Energy µ-amu |
|---|---|---|---|---|---|---|
| C Carbon | 6 | 10 | 10.016,856,4 | +Beta | 64,754 | 3,322 |
| | | 11 | 11.011,433,3 | +Beta | 78,842 | 2,128 |
| | | 12 | 12.000,000,000 | | 98,940 | (-) |
| | | 13 | 13.003,354,826 | | 104,250 | (-) |
| | | 14 | 14.003,241,982 | -Beta | 113,028 | 168 |
| | | 15 | 15.010,599,2 | -Beta | 114,335 | 10,490 |
| | | 16 | 16.014,701 | -Beta | 118,898 | 9,601 |
| N Nitrogen | 7 | 12 | 12.018,613,0 | +Beta | 79,487 | 18,613 |
| | | 13 | 13.005,738,60 | +Beta | 101,026 | 2,384 |
| | | 14 | 14.003,074,002 | | 112,356 | (-) |
| | | 15 | 15.000,108,97 | | 123,986 | (-) |
| | | 16 | 16.005,099,9 | -Beta | 127,660 | 10,185 |
| | | 17 | 17.008,450 | -Beta | 132,974 | 9,319 |
| O Oxygen | 8 | 14 | 14.008,595,33 | +Beta | 105,994 | 5,521 |
| | | 15 | 15.003,065,4 | +Beta | 120,189 | 2,956 |
| | | 16 | 15.994,914,63 | | 143,724 | (-) |
| | | 17 | 16.999,131,2 | | 148,172 | (-) |
| | | 18 | 17.999,160,3 | | 156,808 | (-) |
| | | 19 | 19.003,577 | -Beta | 154,337 | 5,174 |
| | | 20 | 20.004,075,5 | -Beta | 162,504 | 4,094 |

The data show that *SE* is the touchstone of nuclear stability. For each *Z* there is a number of isotopes of successively larger *A*. For any *Z* the isotopes of "medium values of *A*" are stable. They have negative *SE;* that is, the total mass / energy of the nucleus is not large enough to make up any set of decay products whatsoever.

Those of smaller *A* have positive *SE* and emit a particle which in most cases *(+Beta*, a positron) changes the type to being type *[Z-1]* at the same *A*, a step toward being a type "of medium $A$" for the new, lower $Z$ that it has become. (In some cases a different particle is emitted but the tendency to change toward a type where the $A$ is "medium" is always the case.) For example, unstable type $_7N^{12}$ emits a $+Beta$ and becomes stable type $_6C^{12}$.

Likewise, the types of relatively large $A$ for their $Z$ also have positive $SE$. They in most cases emit a particle ($-Beta$, an electron) which changes the type to being type $[Z+1]$ at the same $A$, a step toward being a type "of medium $A$" for the new higher $Z$ that it has become. For example, unstable type $_7N^{17}$ emits a $-Beta$ and becomes stable type $_8O^{17}$.

So to speak, all atomic nuclei are unstable; however, there are no products to which those with negative $SE$ can decay; they are forced into stability by the requirements of conservation of mass / energy. Those with positive $SE$ can and do decay and the process, the nature of the particle emitted, is such as to move them toward being stable types.

Why is this so ? It is the pattern of the nuclear masses that creates this situation. As indicated hypothetically and greatly exaggerated in Figure 2, below, it is the way that the mass varies from isotope to isotope that results in a narrow range of nuclei with negative $SE$ and consequent stability, the nuclei on either side of that range having positive $SE$ and consequent instability. For a given $Z$ the masses of the isotopes are not exactly some constant number times $A$; rather they vary from such a straight line relationship, only very slightly, in an *S-shaped* curve fashion.

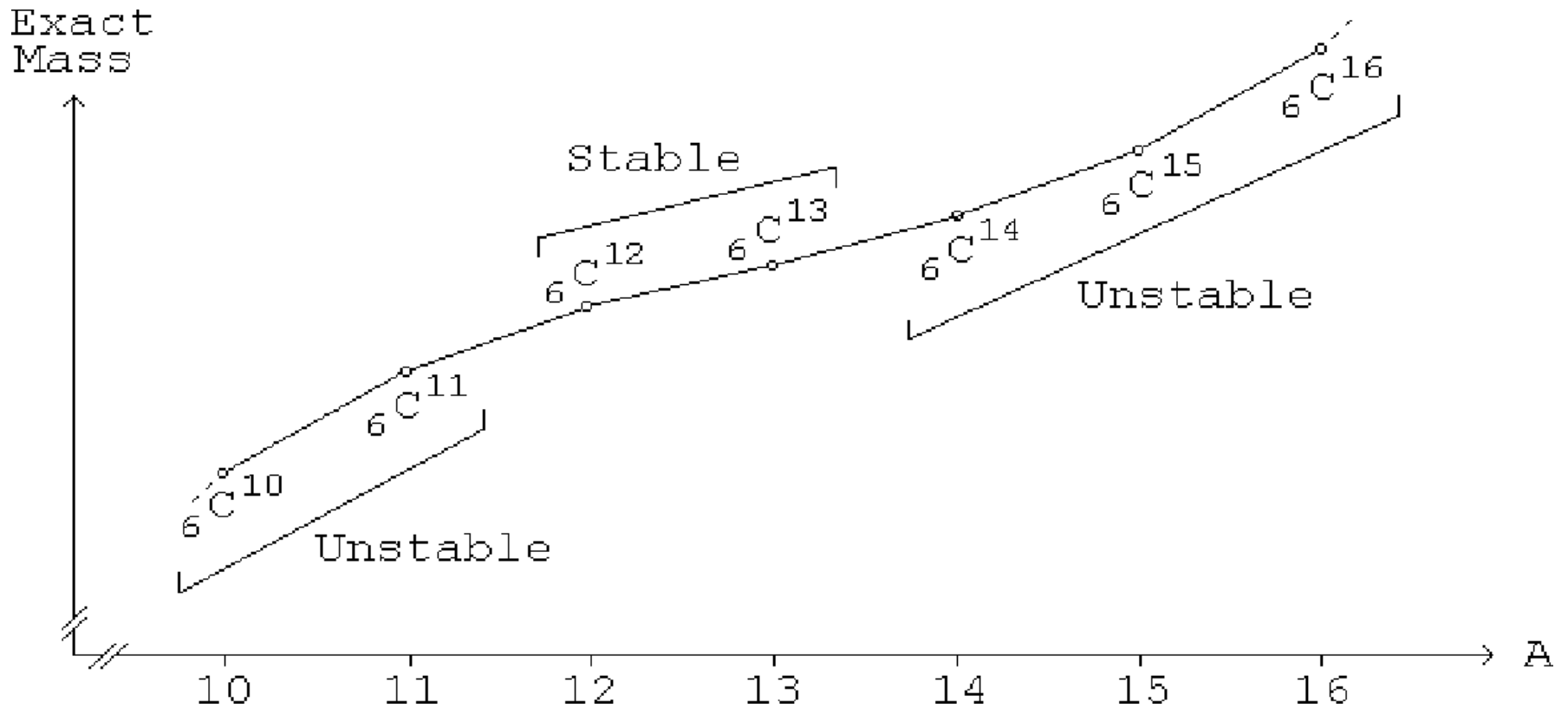

*Figure 2*
*Curvature in Isotope Nuclear Mass Variations*
*(Hypothetical and Exaggerated)*

This curvature in the variation of mass, which is so important and significant, is too small to be observed in a practical unexaggerated plot. If, instead, the plot is modified to $[A - Exact\ Nuclear\ Mass\ (amu)]$ versus $A$ then only the deviations from linearity are plotted, the changes in curvature which range from small to large to small again. Figure 3, on the next two pages, is a precise and accurate plot of that curvature change for selected example nuclear types.

The curves of Figure 3, below demonstrate the curvature of the change in mass from isotope to isotope that results in the *S-shaped* curvature.

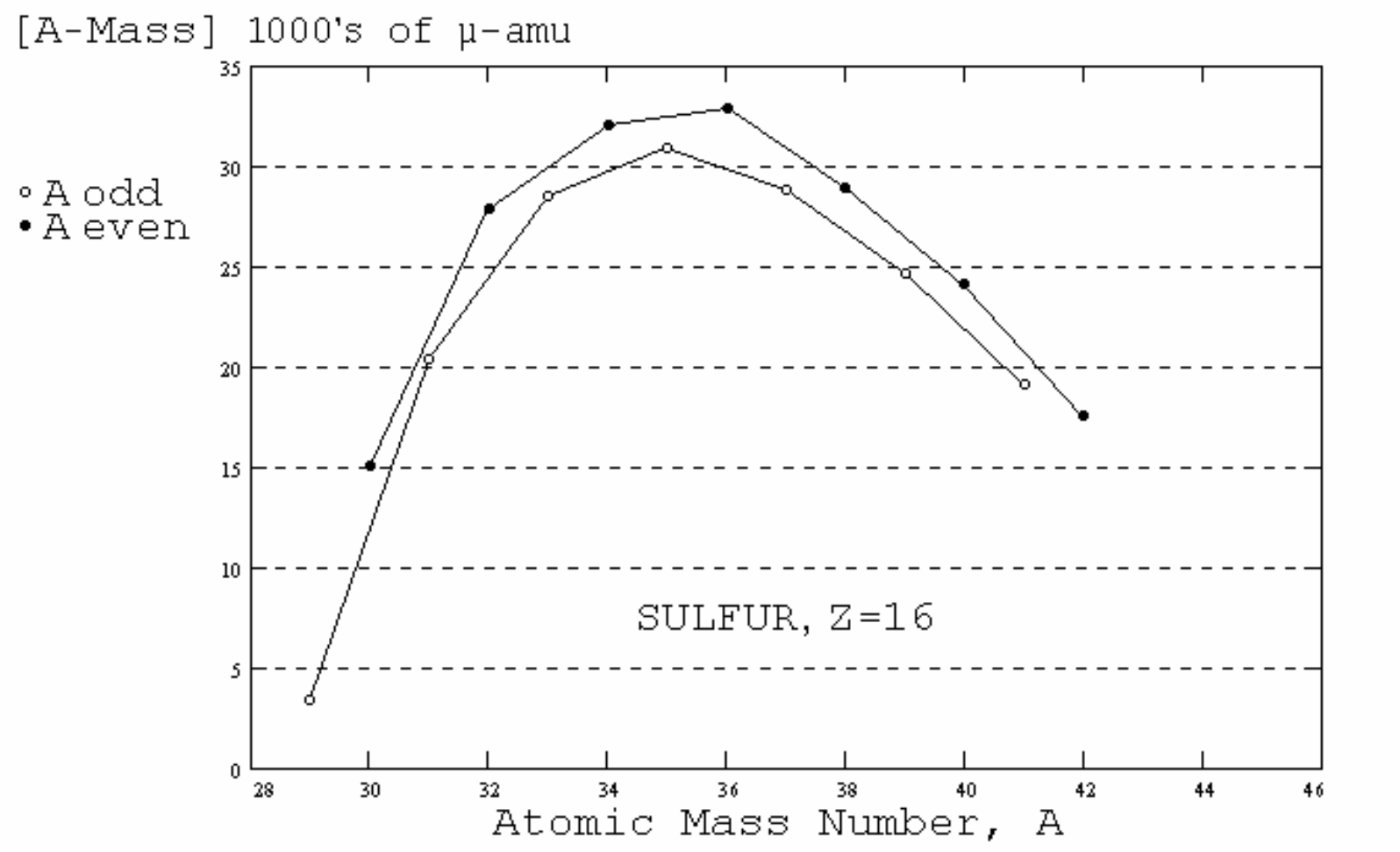

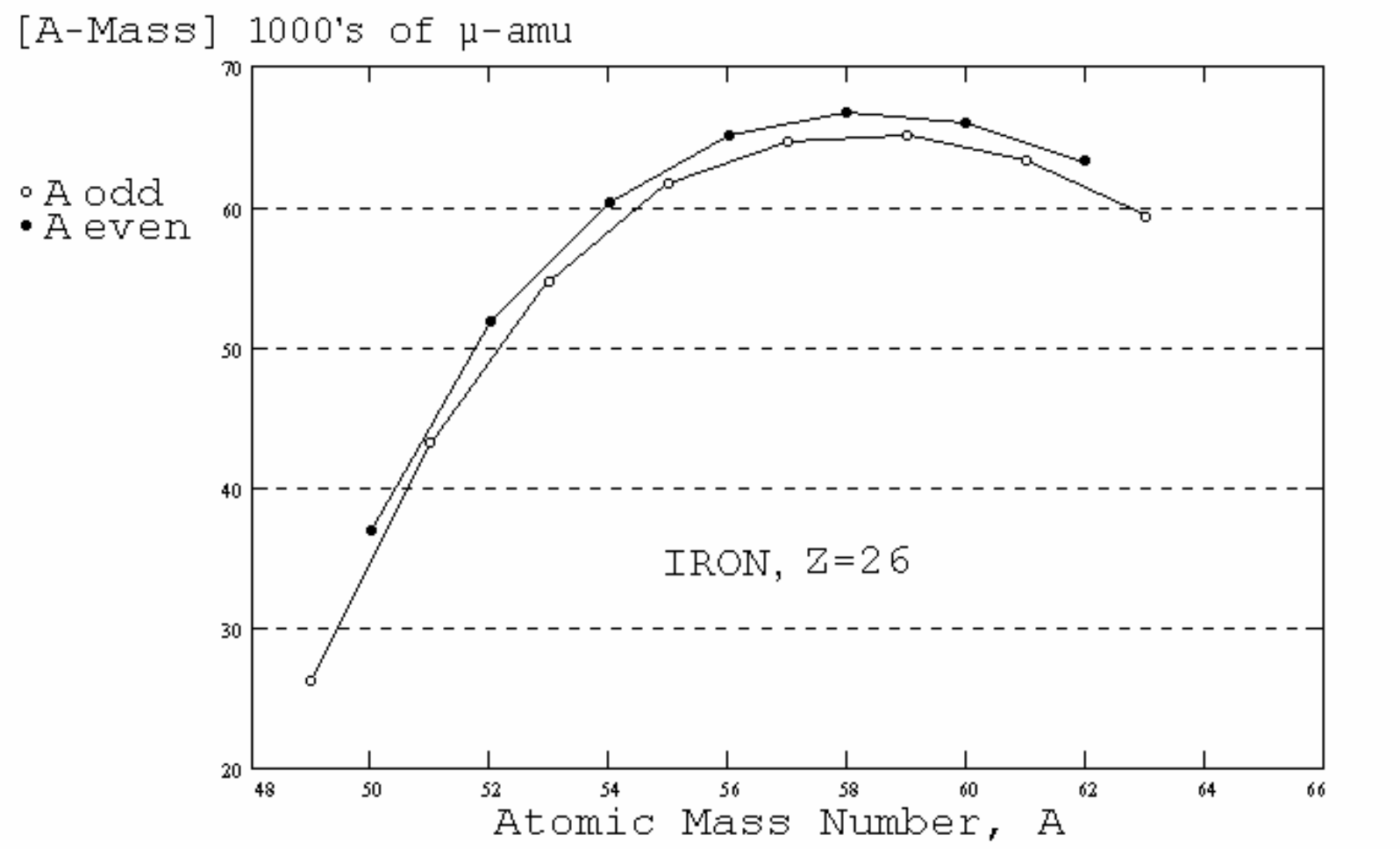

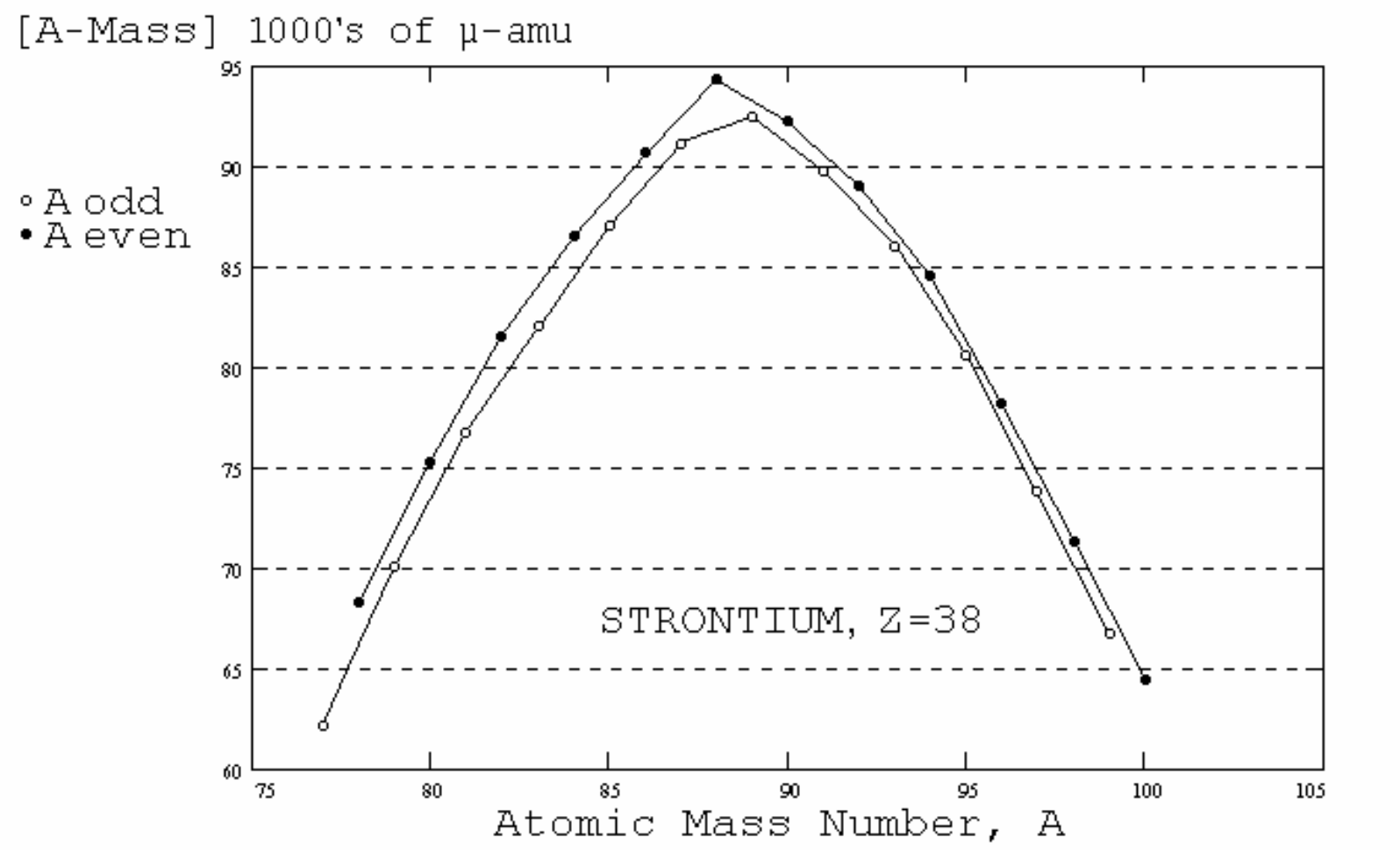

*Figure 3, Page 1*

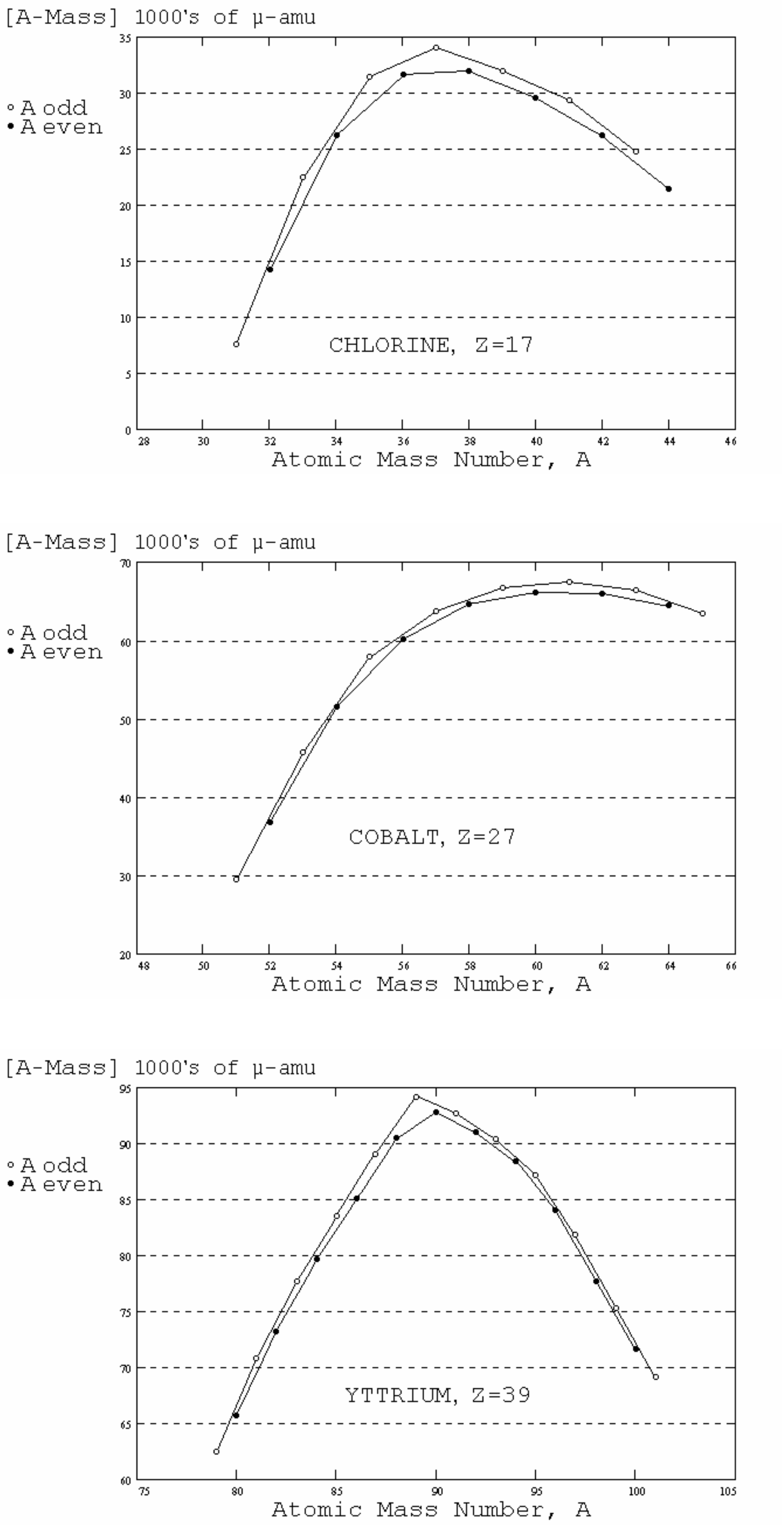

*Figure 3, Page 2*

The curves of Figure 3, in addition to demonstrating the curvature of the change in mass from isotope to isotope, disclose two other phenomena that are important:

(1) There is a distinction in mass variation pattern between isotopes of odd $A$ and those of even $A$. Two separate curves appear for each $Z$, one curve for the odd $A$ isotopes and another for the even $A$ ones.

(2) There is a distinction in the mass variation of odd and even $Z$ types. In odd $Z$ types the odd $A$ isotopes are the upper curve in each plot. For the even $Z$ types the odd $A$ isotopes are the lower curve.

These two data combined would indicate that the distinction is in the number of neutrons, whether odd or even. Since the number of neutrons is $[A - Z]$, then if both $A$ and $Z$ are odd or both even, the number of neutrons will be even. If one of $A$ and $Z$ is odd and the other is even the number of neutrons will be odd. Since the vertical axis in the plots is proportional to actual nuclear mass the curves indicate that nuclei of even $N$ are slightly less massive.

The *The 1983 Atomic Mass Evaluation* data appear to be chaotic in their minor variations, the aspect crucial to the behavior of matter. But, since nature is orderly, there must be an underlying pattern or patterns that account for the exact masses, which are themselves the cause of the overall pattern of stable and unstable types. It is those patterns that must be found and included in the nuclear model.

The trend of $N/A$ versus $A$, Figure 4, below, shows such patterns.

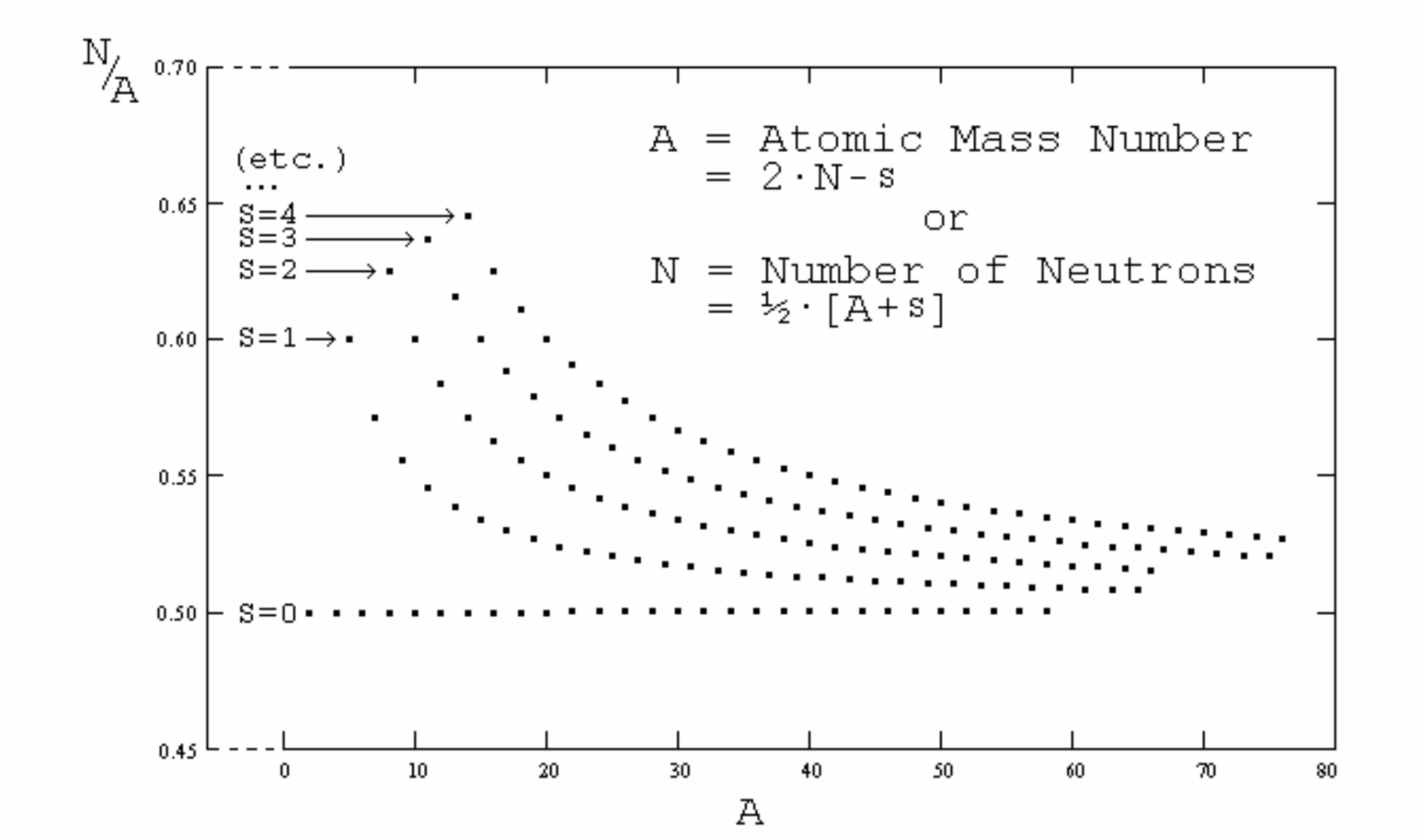

*Figure 4*

The seemingly chaotic pattern of the minor variations in the nuclear masses is here orderly. They appear in series according to the relative amounts of the particles as in equation *(2)*, which suggests a pattern in the structure of the nuclei.

$$A = 2 \cdot N - s \quad [s = \text{series number}] \tag{2}$$

However, Figure 4 is in terms of the integers, $A$ and $N$, not exact masses. That a set of integers produces an orderly pattern does not necessarily mean that the actual exact masses are orderly.

The nuclear masses within an `s-series` appear in Figure 5 on the following page. The same quantity as in Figures 3, is employed except modified to *[A - mass] ÷ A* to accommodate the much greater mass range to be treated. Because of the distinction between odd and even $Z$ types the two are analyzed separately.

The Figure 5 data would appear to indicate that there is a simple and regular mode of behavior, structure or process that operates effectively for high $Z$ or high $s$ series, that is, the variations from nuclear type to type are smooth and regular there. That mode appears to also operate for low $Z$, low $s$ series, but would appear to there be partially overwhelmed by some other effect not so far detected and taken into account.

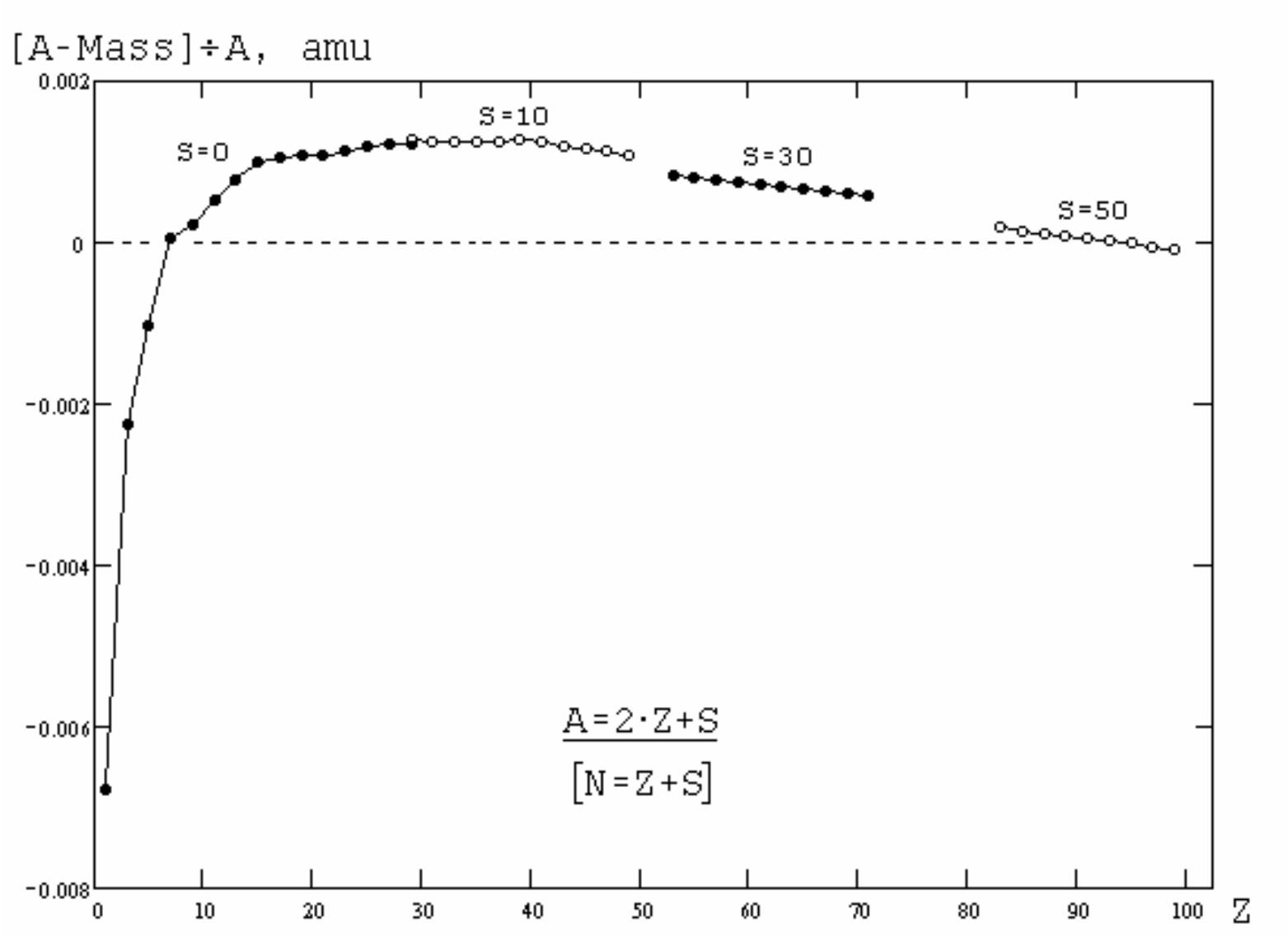

*(a) Odd Z's*

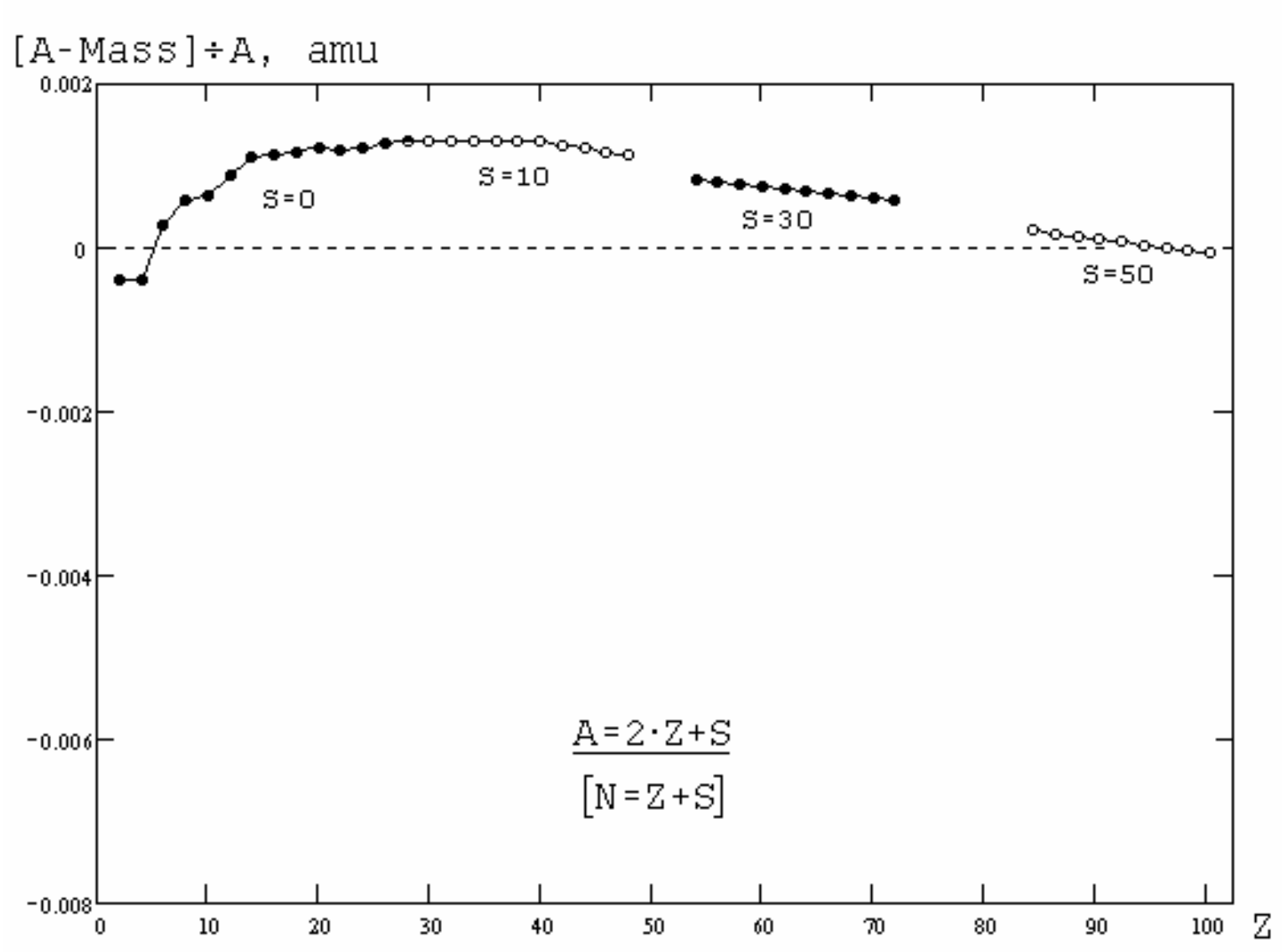

*(b) Even Z's*

*Figure 5*

To analyze the process operating at low $Z$ or on low $s$ series, Figure 6, below, investigates the same changes as Figure 5, but for several adjacent low $s$ series: $s = -1$, $0$, and $+1$. The outstanding characteristics of these data is that regular dips or valleys in the graphs occur at values of $Z$ just following each of $Z = 4$, $8$, and $20$.

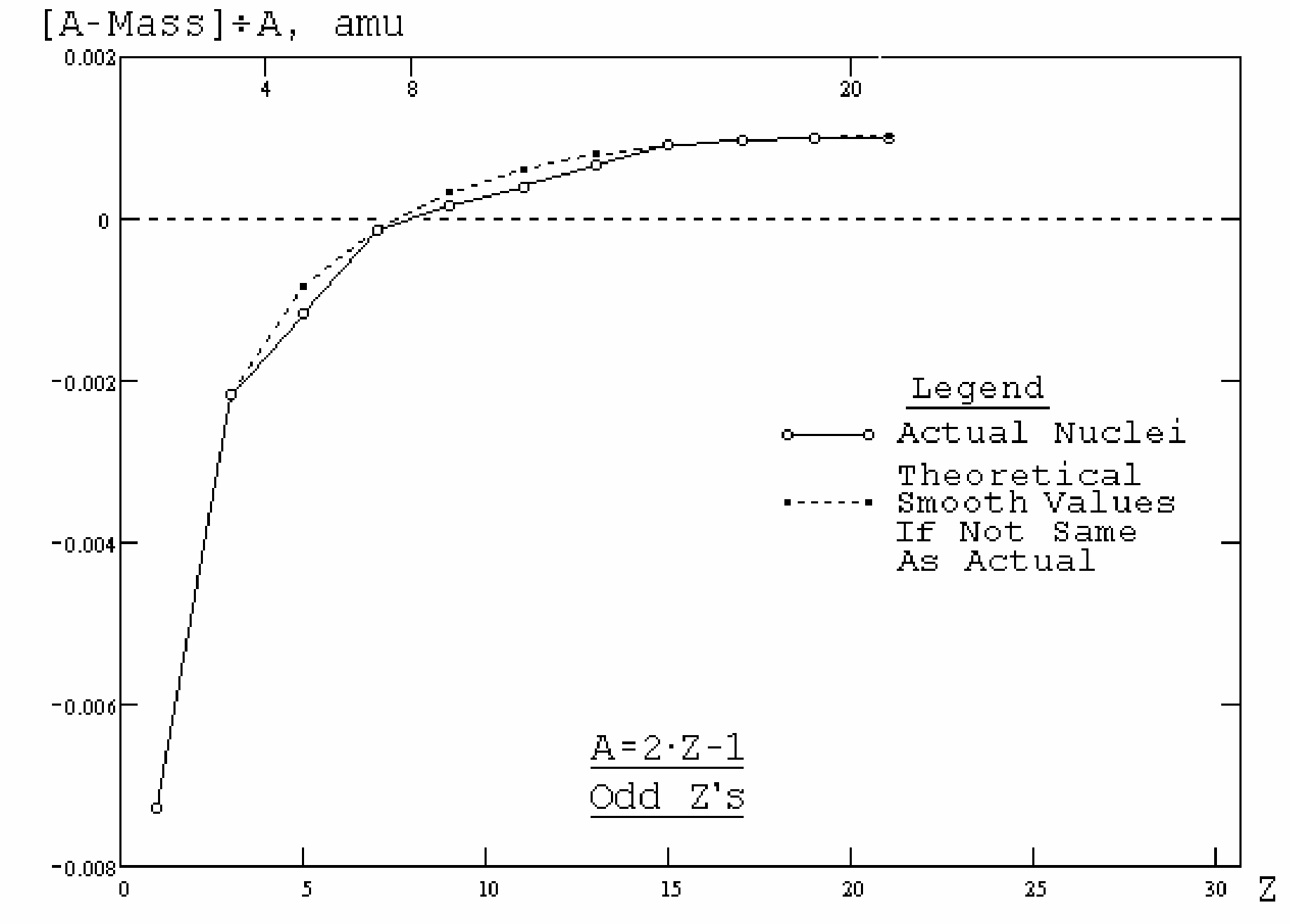

*(a) Odd Z's, Series s = -1*

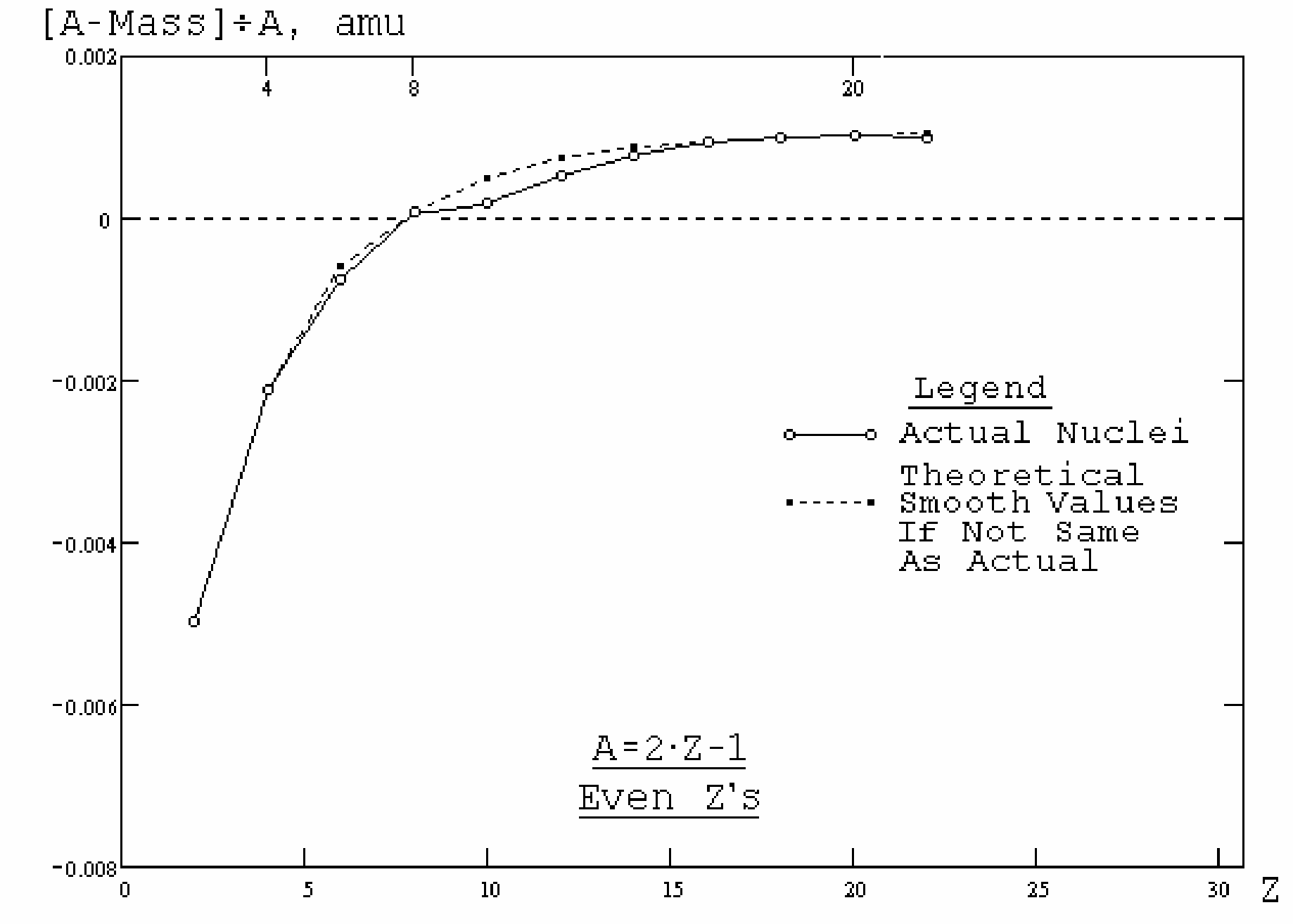

*(b) Even Z's, Series s = -1*

*Figure 6, Page 1*

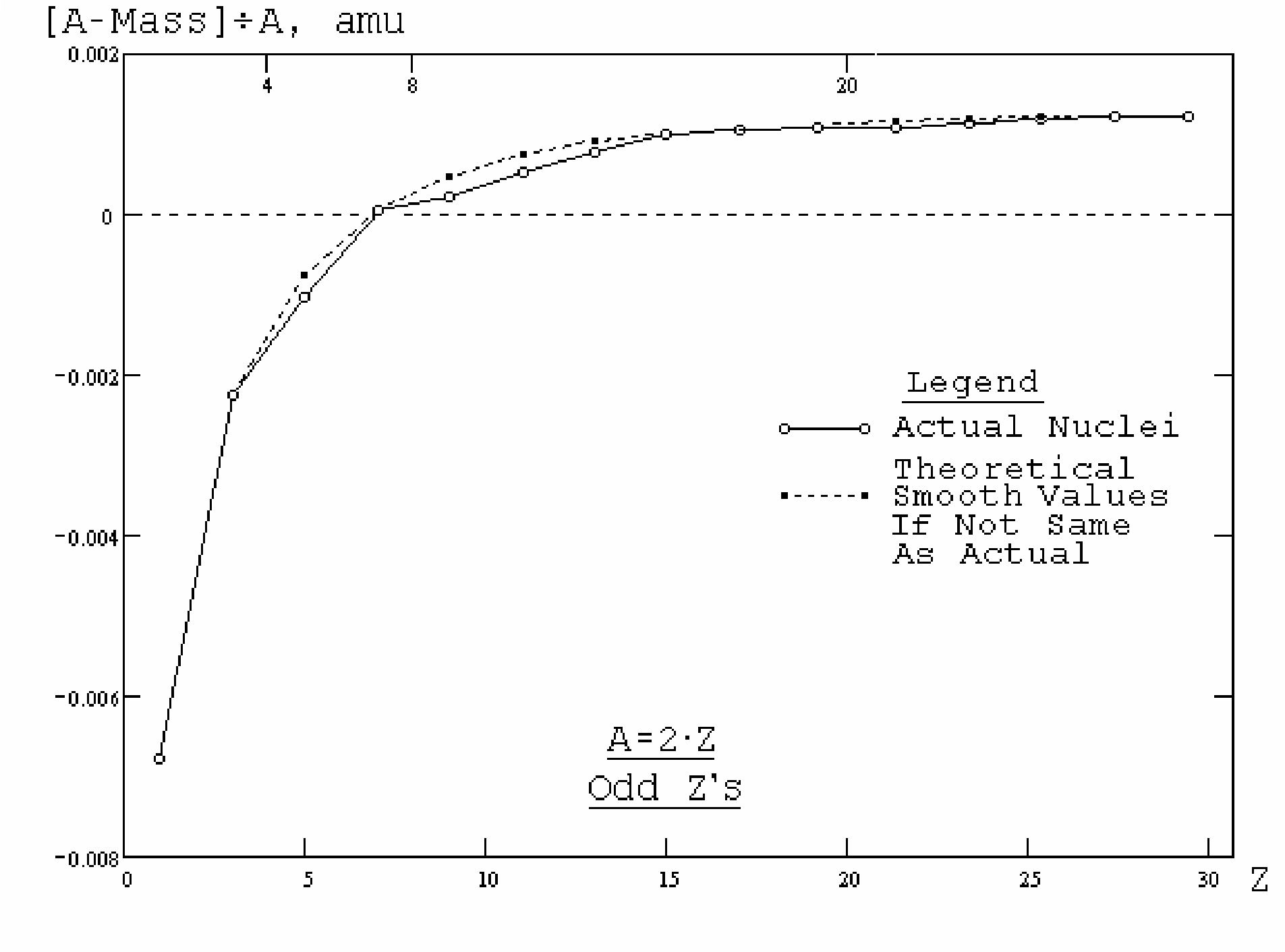

*(c) Odd Z's, Series s = 0*

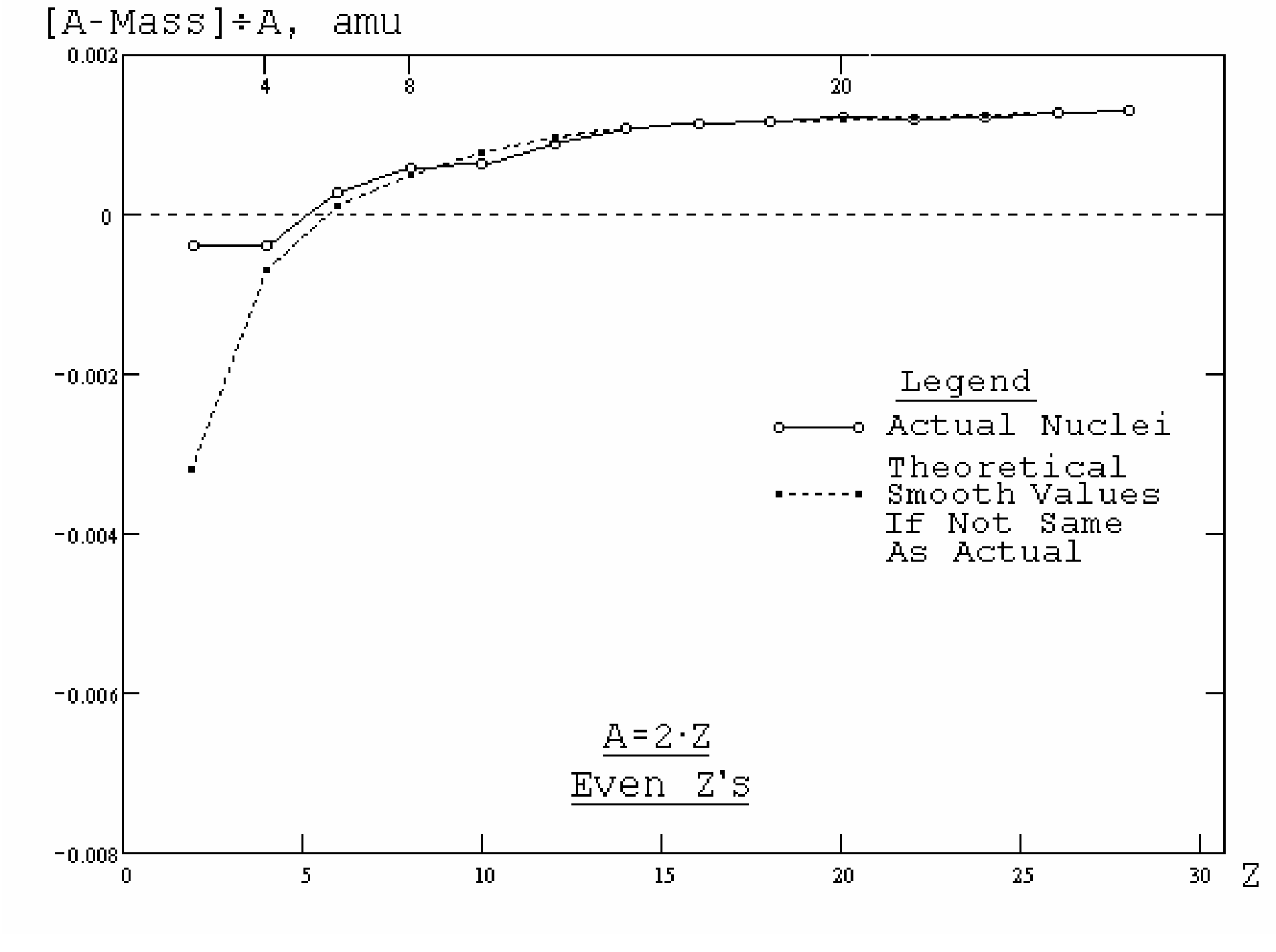

*(d) Even Z's, Series s = 0*

*Figure 6, Page 2*

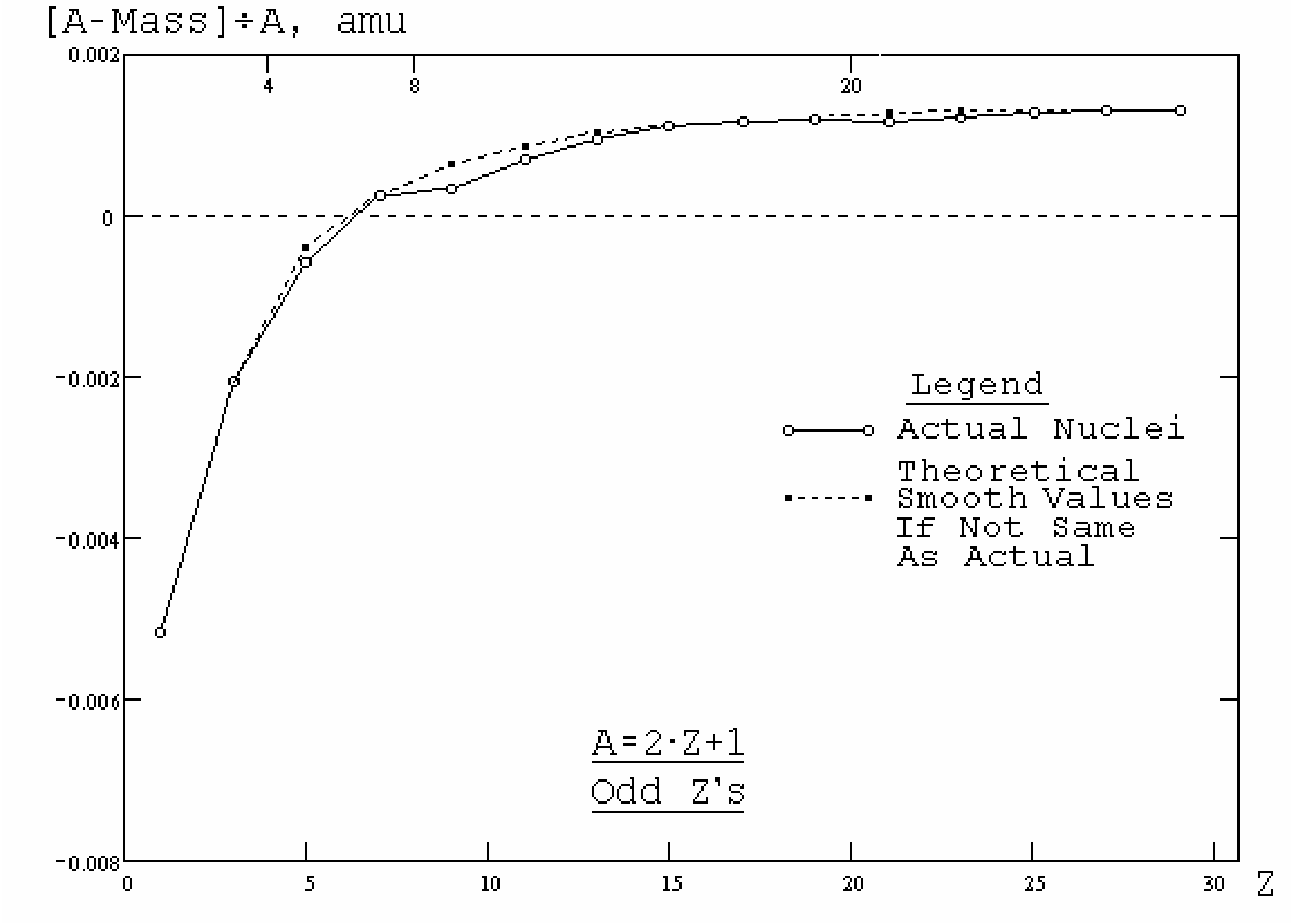

*(e) Even Z's, Series s = +1*

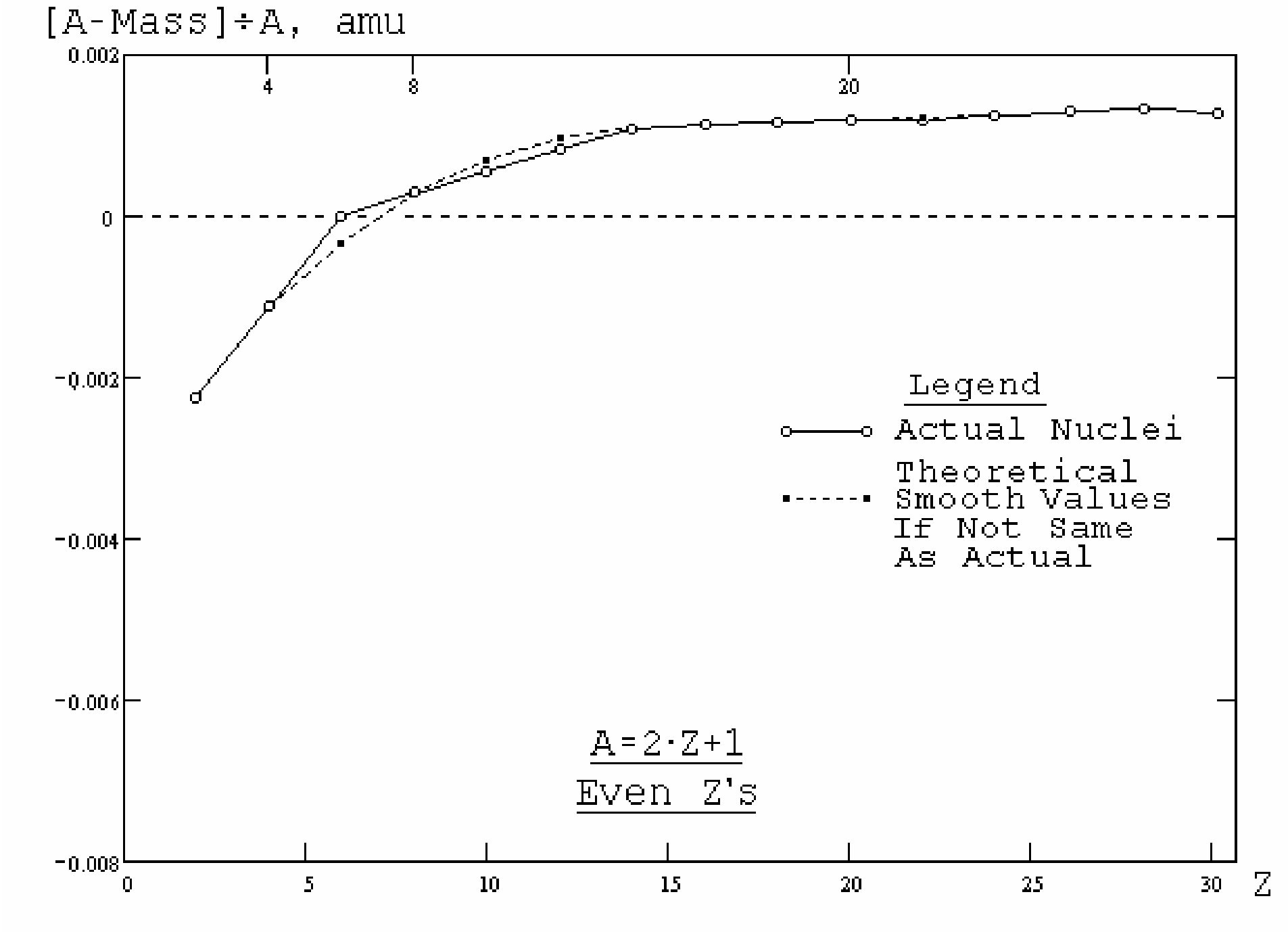

*(f) Even Z's, s = +1*

*Figure 6, Page 3*

In order to understand the effect operating here a brief digression into a relatively slightly attended area of mathematics is necessary. The subject area is that of *polytopes*. A polytope is a geometric figure in $[n]$ dimensions having as its boundary a number of geometric figures in $[n - 1]$ dimensions. If the boundary figures are all identical then the polytope is *regular*, and it is regular polytopes that are of interest here.

A one - dimensional polytope is a simple straight line having as its boundary its zero - dimensional end points (and being not of much interest as a polytope). A two - dimensional polytope is a (flat) polygon, having one - dimensional straight lines as its boundary, examples of regular polygons being the equilateral triangle, the square, and so forth. A three - dimensional polytope is a *polyhedron*. Its boundary is a set of flat *faces* that are polygons. Some common polyhedrons are the pyramid and the cube.

It turns out that the regular polyhedrons are significant in atomic nuclear structure. There are only five regular polyhedrons that can exist (they are sometimes referred to as the *Platonic Solids* because Plato was the first to recognize and study them) and these are listed in Table 7, below.

| Name | Face | Nr Of Faces | Surface Area | Volume | Radius of Inscribed Sphere |
|---|---|---|---|---|---|
| Tetrahedron | Equilateral Triangle | **4** * | $1.73 \cdot a^2$ | $0.12 \cdot a^3$ | $0.20 \cdot a$ |
| Cube | Square | 6 | $6.00 \cdot a^2$ | $1.00 \cdot a^3$ | $0.50 \cdot a$ |
| Octahedron | Equilateral Triangle | **8** * | $3.46 \cdot a^2$ | $0.47 \cdot a^3$ | $0.41 \cdot a$ |
| Dodecahedron | Regular Pentagon | 12 | $20.65 \cdot a^2$ | $7.66 \cdot a^3$ | $1.11 \cdot a$ |
| Icosahedron | Equilateral Triangle | **20** * | $8.66 \cdot a^2$ | $2.18 \cdot a^3$ | $0.75 \cdot a$ |

Where "a" is the length of an edge, one straight line segment of a face's boundary. The values shown are two decimal places of irrational numbers except for the cube.

*Table 7*
*The Regular Polyhedrons*

The appearance in the above table of the same three key numbers: 4, 8, and 20, that turned up in the graphs of Figure 6 is immediately noticeable. But, the major significance is that those three cases have relatively the smallest overall sizes; they are the most compact. That is apparent from the relative volumes, relative surface areas and relative inscribed spheres indicated in Table 7. Figure 8 on the following page depicts these five polyhedrons to the same scale, that is the same edge length, "a" in the above table. The relative compactness of the three equilateral triangle faced polyhedrons is apparent.

The relationship between these solid geometric forms and the atomic nuclear structure, which relationship would appear to be indicated by the correlation of the number of faces of the three most compact of the five regular polyhedrons with the regular dips in the mass curves of the low $Z$, low $s$ atomic nuclear types, is as follows.

(1) The theoretical assembly of an atomic nucleus from its component particles involves the assembling together of a number of like charges: a number, $N$, of electrons and a larger number, $A$, of protons.

(2) In such an assembling of like charges, for example the electrons, the like charges all mutually repel each other with the Coulomb force. Consequently, they would naturally space at equal separation distances in the form of a sphere in space. Assembling them into a nucleus is a case of reducing the size of that sphere to the point where the individual particles essentially merge.

(3) That configuration in space before the merging is geometrically equivalent to the sphere inscribed inside a regular polyhedron -- at least when the number of merging particles is one of the five cases of Table 7. The center of each face of the polyhedron corresponds to the location of the charges. The inscribed sphere touches each face at just that point.

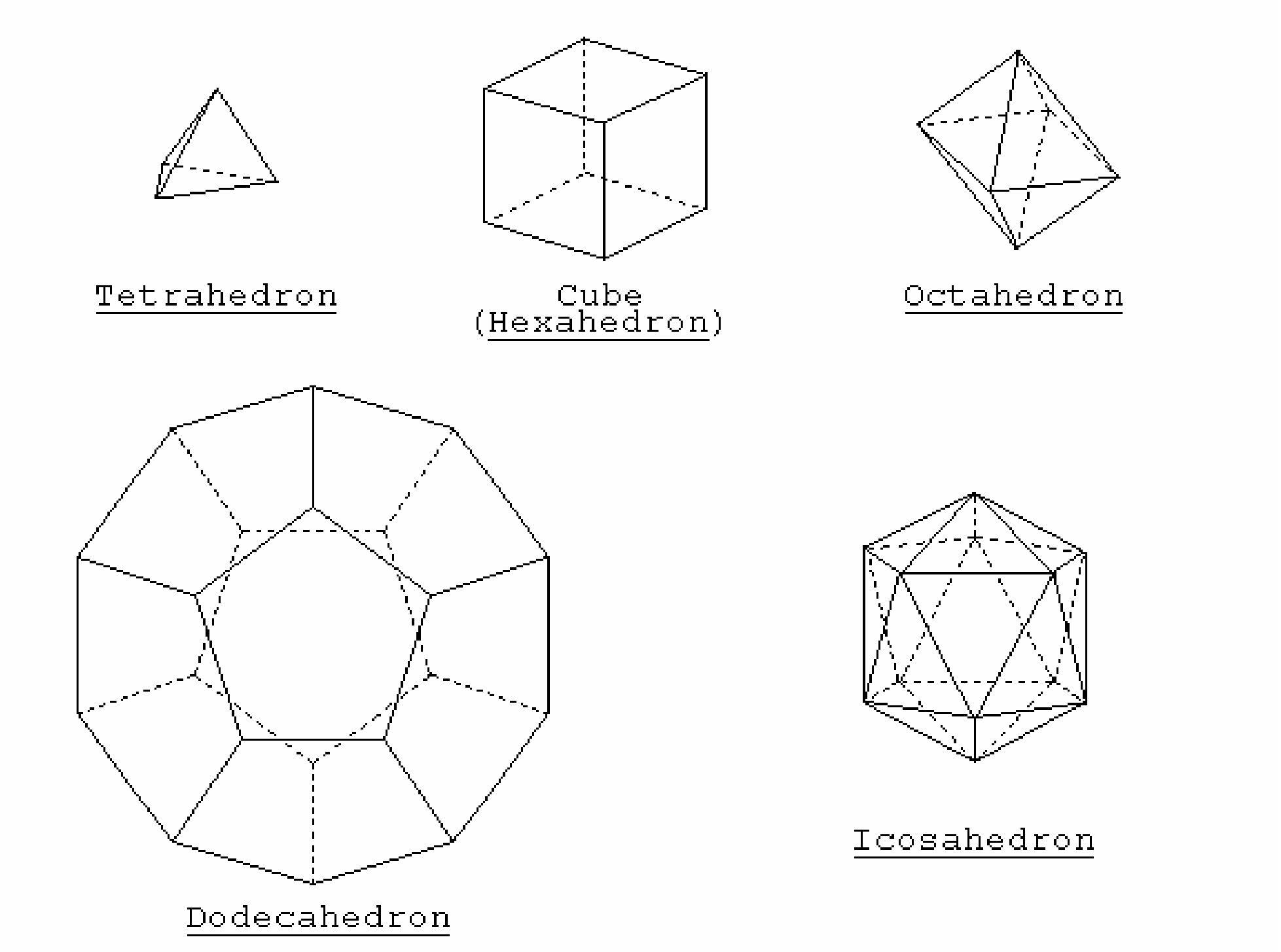

*Figure 8*
*The Regular Polyhedrons*

When the number of merging particles does not correspond to the number of faces in one of the five regular polyhedrons the configuration of the mutually repelling particles is still according to a polyhedron having its number of faces equal to the number of like charge particles that are merging. However, the polyhedron is not regular and that means that the particles are unable to space equally. The best that they can do is arrive at some more or less stable balanced mixture of separation distances that vary around the average value.

The resulting corresponding polyhedron is a quasi-regular form having polygons of various numbers of sides as its faces. It is not as compact as would be the case if it were regular, however. Its inscribed sphere does not touch all of its faces, only the nearest ones, and that means that some of the charges are radially farther from the center than others.

If the polyhedron corresponding to the assembling charges is regular then the radial distance of each of the charges from the center is the same and the configuration is more compact. Furthermore, if the polyhedron is of the type having $4$, $8$, or $20$ faces, representing an assembly of $4$, $8$, or $20$ like charges, then the radial distance of each charge from the center or the group is a minimum; the configuration is maximally compact.

The more compactly that these like charges can fit together the greater will be the potential energy between them and, consequently, the greater will be the energy which must be removed from them for their merging into a new nucleus to take place. Therefore compactness of the natural configuration of the like charge particles assembling into a nucleus corresponds directly to the mass decrease exhibited by that nuclear type.

In the graphs of Figure 6 the vertical axis is $[A - Mass] \div A$. Therefore, smaller mass (greater mass decrease) produces higher points on the curves, larger mass (smaller mass decrease) produces dips in the curve. The high points on the curves correspond to greater compactness of the assembly configuration. The dips correspond to less compact cases.

Now, in the assembling of $N$ electrons and $A$ protons, the $N$ electrons and a corresponding $N$ out of the total of $A$ protons offset each other. Their merger of mutual attraction occurs naturally and readily. Only the excess $Z$ protons remaining have the above described configuration problems as they are being assembled into a nucleus. Thus results the significant points in the curves of Figure 6 at $Z = 4$, $8$, and $20$.

Consider the Odd curves of Figure 6 in the region near $Z = 8$.

> If the $Z = 7$ case is thought of as $Z = 8$ with one proton removed (one not neutralized by an electron to form a neutron, but instead one taking a position according to the polyhedron), then that case of $Z = 7$ is at least as compact as that for $Z = 8$ and actually can be a little more compact because of the missing proton. However the value of $A$ is reduced and the net value of $[A - Mass] \div A$ (the value of which $= 1 - Mass/A$) is somewhat less than that for $Z = 8$. The type thus plots on the graph as on the main trend a little lower than where the $Z = 8$ would fall.
>
> The $Z = 9$ case, if viewed as $Z = 8$ with one excess proton added, certainly must be less compact than $Z = 8$. Being so it has less overall mass decrease than for $Z = 8$, therefore greater total mass and plots at a lower point on the curve, off of the smooth trend.
>
> The $Z = 10$ and successively higher $Z$ cases exhibit behavior similar to $Z = 9$ gradually tapering off as the general moderating effect of increasing $A$ and distance from the special $Z = 8$ case increases.

At $Z = 20$ the general cycle repeats except moderated by the much larger value of $A$.

Considering the Figure 6 Even curves cases the behavior is even more pronounced at $Z = 8$ and $20$ because the even $Z$'s produce data points exactly at those key numbers. Below $Z = 8$ the same effects are operating but with modified results because of the even values of $Z$.

> First, at $Z = 6$ the case of the regular polyhedron the cube enters in. It is less relatively compact than the polyhedrons having equilateral triangles for faces, but it is more relatively compact than the cases for $Z = 7$ or $Z = 5$.
>
> In addition, at $Z = 2$ is another maximally compact case. There can be no polyhedron with only two faces, but the configuration is nevertheless as compact as theoretically conceivable, more compact even than the tetrahedron.
>
> The special cases of $_2He^4$ and $_4Be^8$ experience the maximal compactness of $Z = 2$ and $4$ combined with low values of $A$, which tend to make the effects more pronounced. This is especially so for the $s = 0$ cases. There, for Helium and Beryllium, all three of $A$, $Z$, and $N$ correspond to maximally compact cases: $2$, $4$, and $8$.

The last point is of some significance. It must be emphasized that there is no contention that the nuclear type actually materially forms via the simultaneous combining of $N$ electrons and $A$ protons. There is no mechanism available to produce such an effect except within intensely hot stars, and even there the combinations effected must be of two particles at a time. The coincidence of simultaneity required for combining a greater number of particles at a time is prohibitive. The effect of assembly configuration that has been presented stems from that the net resulting atomic nucleus, those nuclei as they must materially exist, must have masses as if they had been so constituted.

Another interesting and useful result develops from the effect of polyhedral geometry on the assembly structure of nuclear types. Table 7 includes as regular polyhedrons only cases with an even number of faces. There are no regular polyhedrons possible having an odd number of faces. The consequence is that the odd $Z$ nuclear types are slightly less compact, have slightly less reduced mass, have slightly greater relative overall masses, and are somewhat less stable or exhibit fewer stable isotopes than their even $Z$ counterparts. It also accounts for the behavior noted a number of pages earlier, with regard to Figures 3, that even $N$ types are slightly less massive (that is have slightly larger mass decrease) than odd $N$ types.

On page 7 in conjunction with Figure 5 it was stated that:

> "These data would appear to indicate that there is a simple and regular mode of behavior, structure or process that operates effectively for high $Z$ or high $s$ series, that the variations from nuclear type to type are smooth and regular there. That mode appears to also operate for low $Z$, low $s$ series, but is apparently there partially overwhelmed by some other effect not so far detected and taken into account."

That behavior is the assembly configuration effect analyzed and developed above and now "detected and taken into account". Without it the variation in mass from nuclear type to type would be completely smooth and regular.

To further investigate the last contention Figures 9(a) and (b), below, amplify the $s = 10$, $30$, and $50$ series presented earlier.

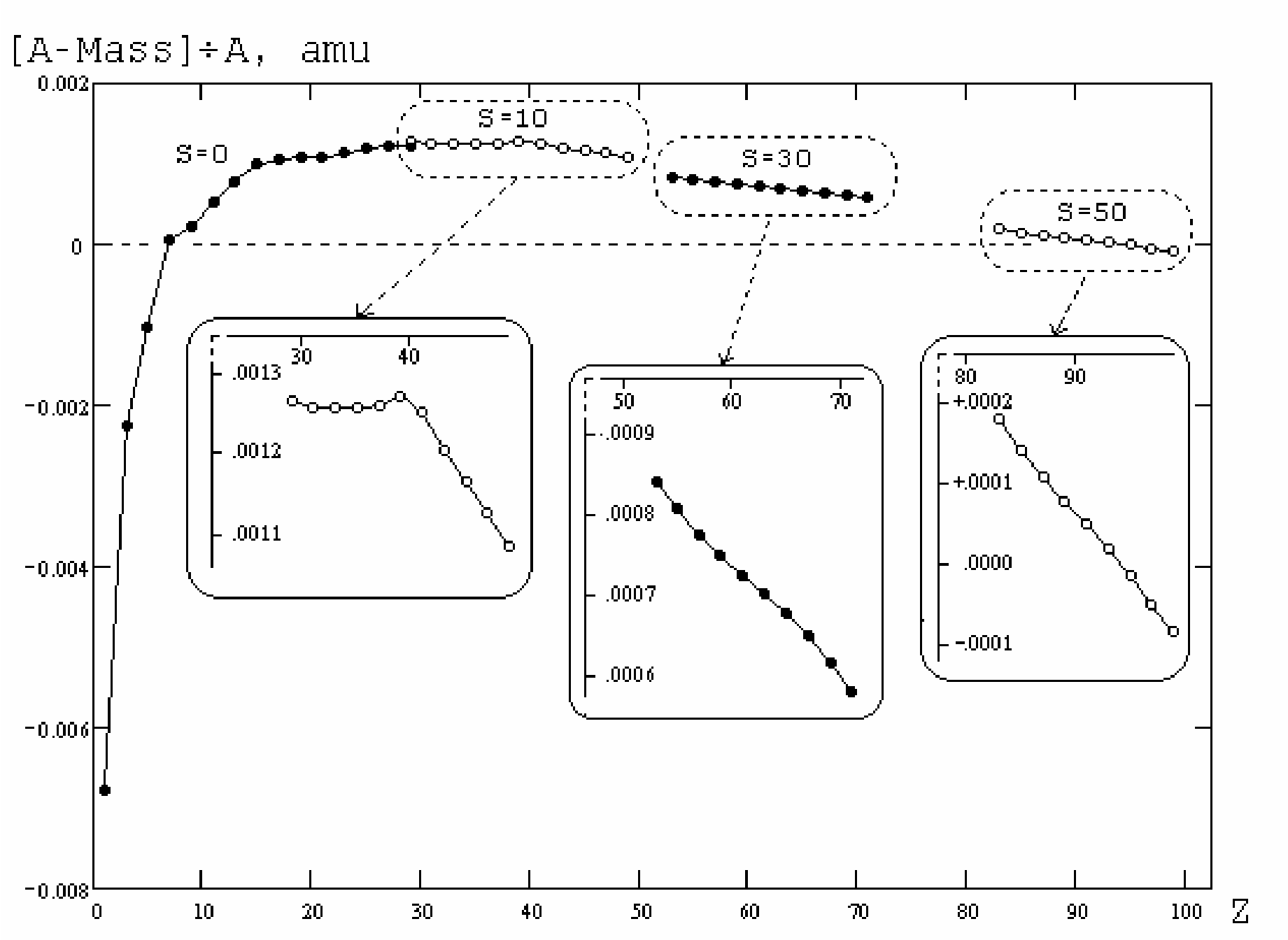

*(a) Odd Z's*

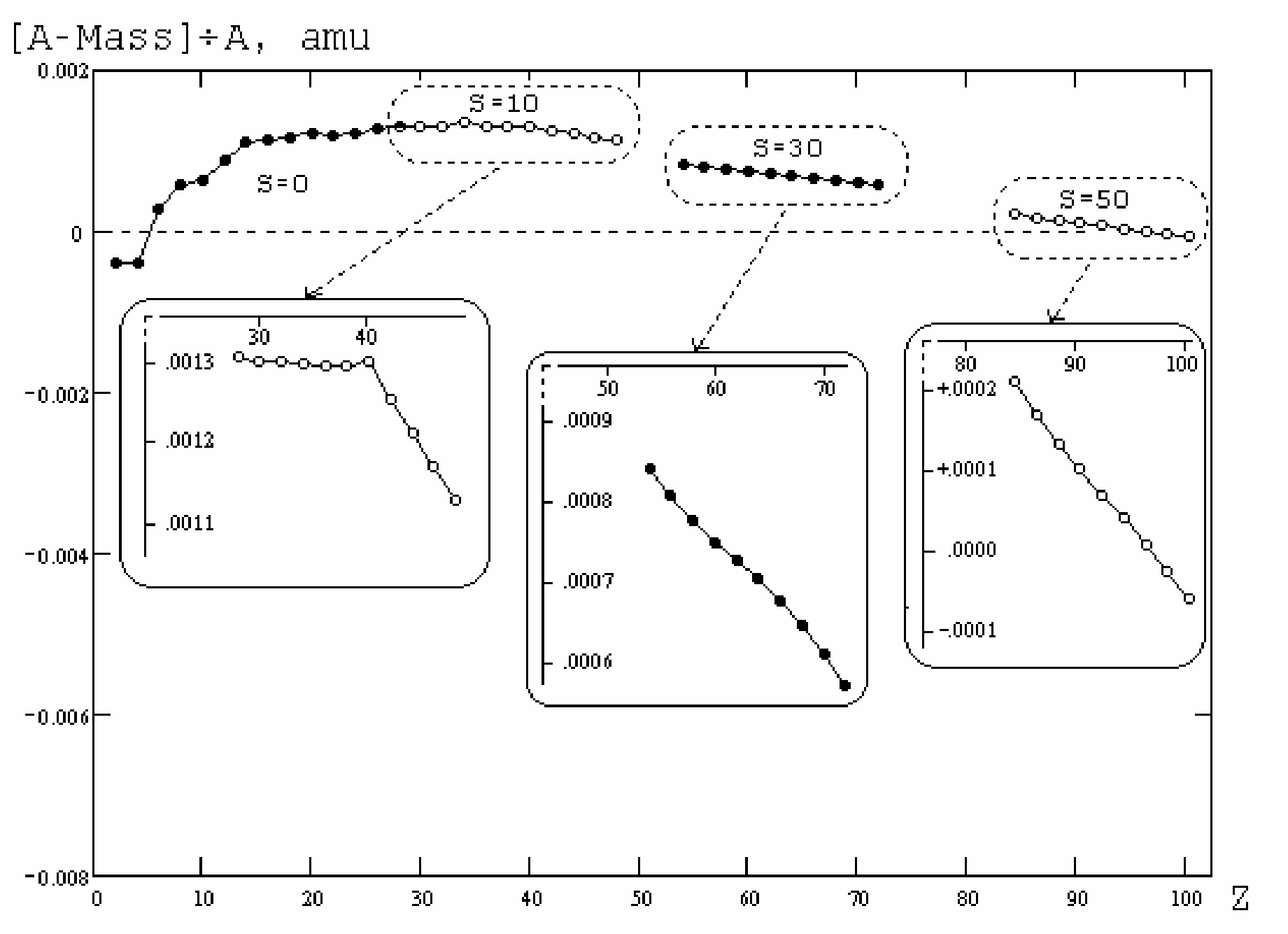

*(b) Even Z's*
*Figure 9*

The series $s = 30$ and $50$ are quite smooth and regular. They exhibit the characteristic *S-type* shape described earlier in conjunction with Figure 2, the characteristic that makes a small number of isotopes in the range at or near $N = A/2$ more stable while all of the other isotopes are unstable. The $s = 10$ series exhibits a minor kink (minor on the original unamplified curve relative to the rest of the curve) just at the $Z = 40$ position (for $Z$ *even*) and (for $Z$ *odd*) just before it.

The minor kink in the $s = 10$ series is due to the polyhedral case of the dodecahedron. If each of the $20$ faces of a dodecahedron is divided in half a quasi-regular $40$ - *faced* polyhedron results. While not as compact as a pure regular polyhedron it is significantly more compact than most values of $Z$ can achieve. It is quite near to being a pure regular polyhedron. Its effect is a "minor kink" because it is not purely regular and because it is at a relatively large value of $A$, which tends to moderate the assembly configuration effect.

While a similar effect might then be expected at $Z = 16$ due to the octahedron, such an effect is of much less significance. Dividing each of the $8$ faces of an octahedron in half is a much greater distortion of the polyhedron than is dividing each of the $20$ faces of the dodecahedron.

The characteristic *S-type* shape, the shape that makes for the stable isotopes amid a sea of unstable ones and, therefore, on which our existence depends, comes about as follows.

> On the one hand, as the number of electrons in the composition of a nucleus becomes greater the number of neutrons becomes greater and, consequently the number of multiples of the $840$ $\mu$-*amu* per neutron mass increase applied to the nuclear type.
>
> On the other hand, as the number of electrons in the composition of a nucleus becomes greater the central negative charge attracting the positive protons as a group becomes larger and tends to produce a more compact overall result.
>
> Thus the first tendency is to increase the nuclear mass and the second is to decrease the nuclear mass, both as $N/A$ increases.

If the $N/A$ ratio is small, that is if there are few electrons in the nuclear composition, then the compactness is quite poor, what with attempting to combine the mutually repelling protons unaided by a central negative charge. If the ratio is large, that is if the nuclear composition is almost all net neutrons, then the neutron mass excesses overwhelm any small mass decrease due to the few un-neutralized protons, even though they are well compacted. Only in the range of balance of these two tendencies can a mass minimum occur. That is at and a little above $N/A = 0.5$.

Figure 10, on the following page is a highly schematic indication of the general form and tendency of those effects.

One other observation concerning these results should be made. For elements of $Z$ higher than $83$, that is *Bismuth*, $_{83}Bi$, there are no stable nuclei at all. The reason for this relates directly to the curvature discussed relative to Figures 2 and 10 and the effect of relative uniformity. Large nuclei vary little in relative composition from isotope to isotope. That is, for large $A$ and large $N$ the ratio $[N]/[A]$ is very little different from the ratio $[N+1]/[A+1]$.

As a result the curvature illustrated in Figure 2, and which effect accounts for a region of negative $SE$ and stability amid unstable surroundings for the lighter types, lessens to the point of ineffectiveness for the heavier types. Apparently the turning point is at *Bismuth*. The amplified depictions of nuclear series in Figures 9 show this effect in that the curvature is quite slight for $s = 50$ compared to that for $s = 30$.

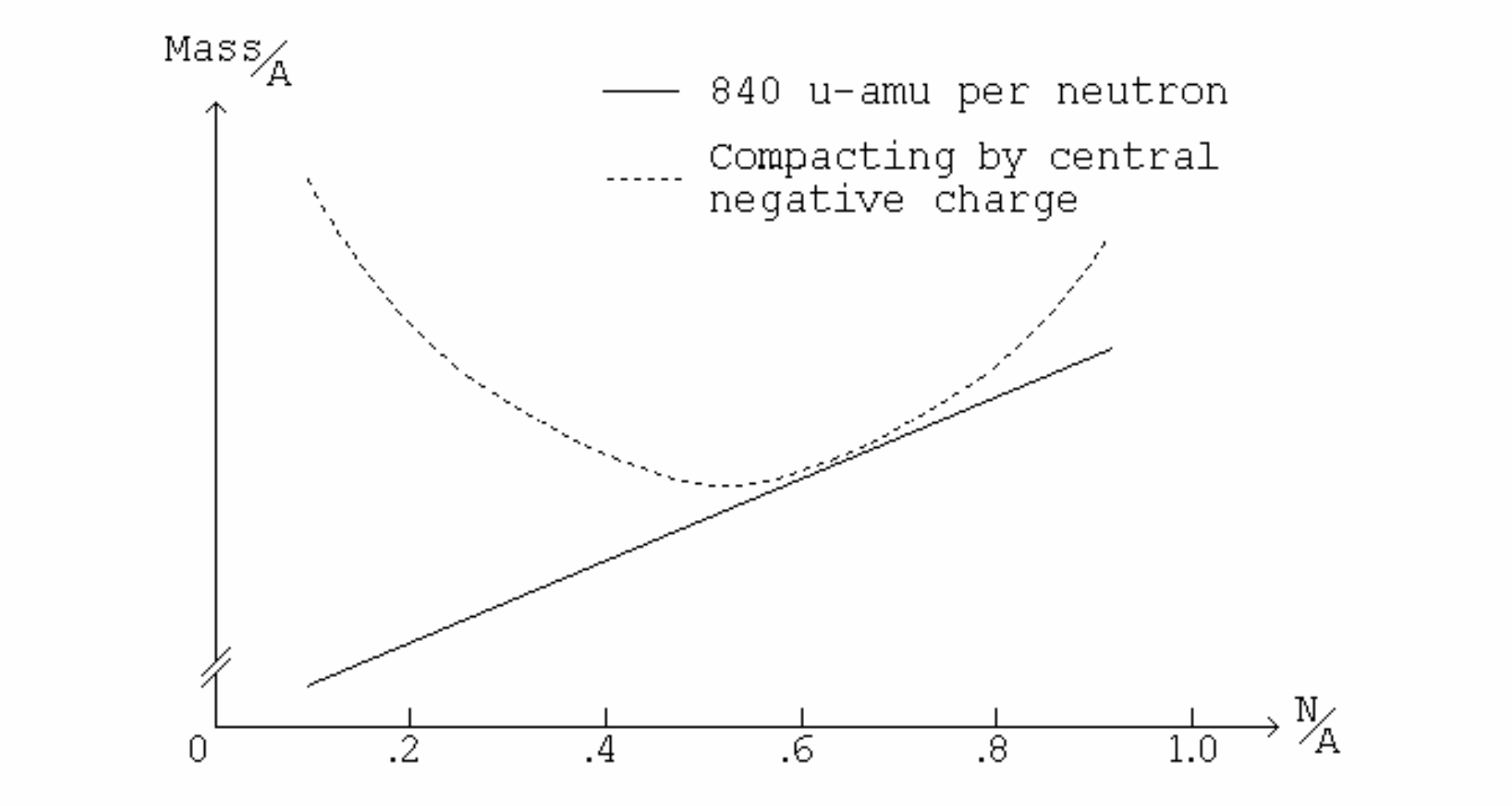

*Figure 10*

Earlier above it was stated as follows:

> "The *The 1983 Atomic Mass Evaluation* data appear to be chaotic in their minor variations, the aspect crucial to the behavior of matter. But, since nature is orderly, there must be an underlying pattern or patterns that account for the exact masses, which are themselves the cause of the overall pattern of stable and unstable types. It is those patterns that must be found and included in the nuclear model."

And, it is those patterns that have been found, developed, and included in the nuclear model in the analyses in the foregoing pages.